\documentclass[pra,onecolumn,epsfig,superscriptaddress,showpacs]{revtex4-2}
\usepackage{graphicx}
\usepackage{subcaption}
\makeatletter
\long\def\@makecaption#1#2{%
  \vskip\abovecaptionskip
  \sbox\@tempboxa{#1: #2}%
  \ifdim \wd\@tempboxa >\hsize
    #1: #2\par
  \else
    \global \@minipagefalse
    \hb@xt@\hsize{\hfil\box\@tempboxa\hfil}%
  \fi
  \vskip\belowcaptionskip}
\makeatother
\usepackage{epstopdf}
\usepackage{amsfonts}
\usepackage{amssymb}
\usepackage{amsmath}
\usepackage{prettyref}
\usepackage{float}
\usepackage{amsmath}
\usepackage{amssymb}
\usepackage{esint}
\usepackage{qcircuit}
\usepackage{color}
\usepackage{CJK}

\usepackage{braket}
\usepackage[left=1in, right=1in, top=1in, bottom=1in]{geometry}
\usepackage{tikz}
\usetikzlibrary{shapes.geometric, arrows}
\def\be{\begin{equation}}
\def\ee{\end{equation}}
\def\ba{\begin{eqnarray}}
\def\ea{\end{eqnarray}}

\def\pnp{${\text P}^{\sharp \text{P}}$ }

\begin{document}
\begin{CJK*}{UTF8}{gbsn}
\title
{Non-Hermitian Computers Need No Complex Numbers}

\author{Qi Zhang(张起)}
\affiliation{College of Science, Liaoning Petrochemical University,
Fushun 113001, China}

%\affiliation{Liaoning Provincial Key Laboratory of Novel Micro-Nano Functional Materials, Fushun 113001, China}

%\author{Biao Wu(吴飙)}
%\affiliation{International Center for Quantum Materials, Peking University, 100871, Beijing, China}
%\affiliation{Collaborative Innovation Center of Quantum Matter, Beijing 100871, China}
%\affiliation{Wilczek Quantum Center, Shanghai Institute for Advanced Studies, Shanghai 201315, China}
%\affiliation{Hefei National Laboratory, Hefei 230088, China}

\begin{abstract}
In traditional quantum computing, it has been established that real quantum computation augmented with non-Clifford gates is as powerful as universal quantum computation. Here we investigate this phenomenon in the non-Hermitian setting. We show that a non-Hermitian quantum computer equipped with the real gate set ${H, \text{CCNOT}, G}$, where $G = \operatorname{diag}(g^{-1}, g)$ with $g > 0$ and $g \neq 1$, can solve problems in $\text{P}^{\sharp\text{P}}$ in polynomial time, matching the capability of its universal non-Hermitian counterpart ${H, T, \text{CNOT}, G}$. This demonstrates that non-unitarity, rather than universality, is the essential resource, and that complex numbers are unnecessary.
\end{abstract}
%\date{\today}
%\pacs{03.65.-w, 03.65.Vf,45.20.Jj}
%03.65.Vf: Phases: geometric; dynamic or topological
%Quantum mechanics, 03.65.-w
%45.20.Jj: Lagrangian and Hamiltonian mechanics

\maketitle
%\newpage
%\large

%\setlength{\baselineskip}{12pt}
\section{Introduction}
Although quantum computers have so far demonstrated only a limited number of concrete advantages over classical computers—such as Shor's algorithm~\cite{Shor} and Grover's algorithm~\cite{Grover}—these examples have motivated research in two complementary directions. One seeks to enhance quantum computational power through hypothetical constructs like closed timelike curves~\cite{Deutsch,Bacon}, nonlinear gates~\cite{Abrams}, postselection~\cite{aaronson,knill}, or Lorentzian gates~\cite{He,ZhangWu}. The other investigates the fundamental sources of quantum advantage, focusing on the constraints under which universal quantum computation reduces to classical simulability.

To explore the fundamental sources of quantum advantage, consider the universal gate set ${H, \text{CNOT}, T}$~\cite{Elementary1995,Nielson}. The Gottesman–Knill theorem shows that restricting to Clifford circuits (comprising $H$, CNOT, and $S = T^2$) enables efficient classical simulation~\cite{Gottesman}, raising the question: can a restricted non-Clifford gate set still retain the full power of universal quantum computation? Fortunately, it has been shown using ancilla qubits that a real quantum computer built from the gate set ${H, \text{CCNOT}}$—where both the gates and the probability amplitudes are real—is exactly as powerful as the universal gate set ${H, \text{CNOT}, T}$~\cite{Shi,Aharonov}. This demonstrates that quantum superposition and entanglement, rather than complex numbers, are the essential sources of quantum advantage.

In the direction of enhancing quantum power, non-Hermitian quantum computers (NQC) offer a striking example~\cite{zhangwu,Barch}. Adding a single non-unitary gate $G=\operatorname{diag}(g^{-1},g)$ (real $g>0$, $g\neq1$) to the universal set gives ${H,T,\text{CNOT},G}$, which dramatically increases computational capability. The class of problems solvable in polynomial time by such devices, denoted BNQP (bounded-error non-Hermitian quantum polynomial time)~\cite{zhangwu}, has been proven to equal $\text{P}^{\sharp\text{P}}$—a sharp contrast to the Hermitian case—although physical resources scale exponentially.

In this work, we adopt a purely computational perspective, bridging the search for fundamental sources of quantum advantage with the pursuit of enhanced computational power. We show that, as in the traditional quantum setting, non-Hermitian quantum computers (NQCs) retain their enhanced capability even when restricted to real gates. Specifically, circuits over the gate set ${H, \text{CCNOT}, G}$—denoted $\mathbb{R}$-NQC (for “real” NQC)—omit the $T$ and $S$ gates and are not universal, yet they preserve the full computational power of the non-Hermitian model. Defining BRNQP (bounded-error real non-Hermitian quantum polynomial time) as the corresponding complexity class, we prove that $\text{BRNQP} = \text{BNQP} = \text{P}^{\sharp\text{P}}$. This demonstrates that non-unitarity, rather than universality, is the essential resource.

Given the recent experimental realizations of non-Hermitian mechanics in platforms such as optical systems~\cite{Zhaoscience,XuNatNano,FengNaturePhoton}, quantum walks~\cite{WangLaser}, cold atoms~\cite{ZhouPRA}, circuit QED~\cite{Starkov,Huang}, PT-symmetric acoustics~\cite{Zhu}, trapped ions~\cite{CaoPRL,Ding}, graphene metamaterials~\cite{LiuOE}, and time-mode-locked lasers~\cite{LeefmansNP}, NQC presents a promising avenue for further investigation.

\section{Power of Purely Real NQC Algorithm }

We have previously shown that the class of problems efficiently solvable by non-Hermitian quantum computation, denoted BNQP, coincides with $\text{P}^{\sharp\text{P}}$~\cite{zhangwu}. (For a formal definition of BNQP, see Ref.~\cite{zhangwu}.) In this section, we present a purely real non-Hermitian algorithm—termed an $\mathbb{R}$ NQC—that also solves a $\text{P}^{\sharp\text{P}}$-complete problem, using only real gates and avoiding the full universal gate set.

First, recall the universal gate set for non-Hermitian quantum computing, which relaxes the unitarity restriction: ${H, T, \text{CNOT}, G}$. The single-qubit gates are the Hadamard $H$, the $\pi/8$ gate $T$, and the non-unitary gate $G$, given by
\begin{equation}
H = \frac{1}{\sqrt{2}}(X + Z),\quad
T = e^{-i\pi/8}\left(\begin{array}{cc} e^{i\pi/8} & 0 \\ 0 & e^{-i\pi/8} \end{array}\right),\quad
G = \left(\begin{array}{lc} g^{-1} & 0 \\ 0 & g \end{array}\right),
\end{equation}
with $g>0$, $g\neq1$. The CNOT gate applies a flip to the target qubit $|\phi\rangle$ iff the control qubit $|\psi\rangle$ is in $|1\rangle$. Together, these gates have been proven sufficient to approximate any transformation without the constraint of unitarity to arbitrary precision~\cite{zhangwu}.

Here we consider a non-Hermitian quantum algorithm for a PP-complete problem (solving any PP problem in polynomial time reduces to solving a PP-complete one). Specifically, we target the MAJSAT problem, defined as: given a Boolean formula $f(x_1,\dots,x_n)$ on $n$ variables, determine whether the majority of assignments satisfy $f=1$, i.e., whether $s > 2^{n-1}$, where $s$ is the number of satisfying assignments. MAJSAT is known to be PP-complete (see, e.g., Ref.~\cite{Arora}).

Crucially, our algorithm does not require the full universal gate set ${H, T, \text{CNOT}, G}$; instead, it relies solely on real gates: the Hadamard gate $H$, the non-unitary gate $G$, and the CCNOT gate—the Toffoli gate with two controls that flips the target if and only if both controls are in the state $|1\rangle$. The set ${H, \text{CCNOT}, G}$ clearly implements only real operations. Although $H$, $T$, and CNOT can generate the CCNOT gate, the set ${H, Z, \text{CCNOT}, G}$ cannot generate $T$ and is therefore not universal.

To formalize this restricted model, we modify the definition of BNQP from Ref.~\cite{zhangwu} by replacing the universal non-unitary gate set ${H, T, \text{CNOT}, G}$ with the restricted set ${H, \text{CCNOT}, G}$. This yields a new complexity class, which we denote as BRNQP. Our aim is to show that $\text{BNQP} = \text{BRNQP}$ by designing an algorithm for MAJSAT within this restricted framework. First, we demonstrate that this restricted gate set is sufficient to solve MAJSAT efficiently. It is worth emphasizing that neither CCNOT nor $G$ belongs to the Clifford group; consequently, the algorithm remains beyond the reach of efficient classical simulation.

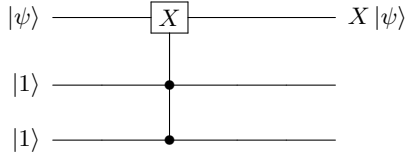
\begin{figure}[htbp]
\centering \hspace{0.0cm}\vspace{0.0cm}
\Qcircuit @C=2em @R=2em {
        \lstick{\ket{\psi}} & \qw & \gate{X} & \qw & \qw & \rstick{X\ket{\psi}} \qw \\
        \lstick{\ket{1}}    & \qw & \ctrl{-1} & \qw & \qw & \qw \\
        \lstick{\ket{1}}    & \qw & \ctrl{-2} & \qw & \qw & \qw
    }
    \vspace{0.5cm}
\caption{The $X$ gate can be implemented using a CCNOT gate with two ancillary qubits prepared in the state $|1\rangle$.}
\label{fig:circuit2}
\end{figure}

In our algorithm, the $X$ gate is frequently required. It can be implemented using a CCNOT gate with two ancillary qubits permanently initialized to $|1\rangle$, as depicted in Fig.~\ref{fig:circuit2}. Moreover, since the Pauli-$Z$ gate satisfies $Z = HXH$, this construction also directly realizes the $Z$ gate. Henceforth, to keep the circuit diagrams uncluttered, we employ the standard symbols for $X$ and $Z$ gates, omitting the ancillary qubits and CCNOT gates used in their implementation.

\subsection{preliminary: determining Whether $f=1$ or $f=0$}

\begin{figure}[t]
\centering
\begin{subfigure}[b]{0.45\columnwidth}
\centering \hspace{0.8cm}
\Qcircuit @C=3em @R=3em {
    \lstick{|c_1\rangle} & \ctrl{3} & \qw & \\
    \lstick{|c_2\rangle} & \ctrl{2} & \qw & \\
    \lstick{|c_3\rangle} & \qw & \ctrl{2} & \qw \\
    \lstick{|a\rangle} & \targ & \ctrl{1} & \qw \\
    \lstick{|t\rangle} & \qw & \targ & \qw
}
\caption{}
\label{fig:ccnot}
\end{subfigure}%
\hfill
\begin{subfigure}[b]{0.45\columnwidth}
\centering \hspace{0.8cm}
\Qcircuit @C=1em @R=1em {
    \lstick{|c_1\rangle} & \ctrl{6} & \qw & \qw &\cdots&  & \qw & \qw & \qw   \\
    \lstick{|c_2\rangle} & \ctrl{5} & \qw & \qw &\cdots & &\qw & \qw & \qw   \\
    \lstick{|c_3\rangle} & \qw & \ctrl{4} & \qw &\cdots & &\qw & \qw & \qw  \\
    \lstick{}   &    &  &     &  \lstick{\vdots~~}  &\rstick{\vdots~~~}& &\lstick{~~~\vdots} & &\\
    \lstick{|c_{n-1}\rangle} & \qw & \qw & \qw &\cdots & &\ctrl{5} & \qw & \qw  \\
    \lstick{|c_n\rangle} & \qw & \qw & \qw &\cdots & &\qw & \ctrl{5} & \qw  \\
    \lstick{|a_1\rangle } & \targ & \ctrl{1} & \qw &\cdots & & \qw & \qw & \qw  \\
    \lstick{|a_2\rangle } & \qw & \targ & \qw &\cdots & & \qw & \qw & \qw  \\
    \lstick{}   &    &  &     &  \lstick{\vdots~~}  &\rstick{\vdots~~~}& &\lstick{~~~\vdots} & &\\
    \lstick{|a_{n-3}\rangle } & \qw & \qw & \qw &\cdots & & \ctrl{1} & \qw & \qw  \\
    \lstick{|a_{n-2}\rangle } & \qw & \qw & \qw &\cdots & & \targ & \ctrl{1} & \qw  \\
    \lstick{|t\rangle} & \qw & \qw & \qw &\cdots & & \qw & \targ & \qw
}
\caption{}
\label{fig:ncnot}
\end{subfigure}
\caption{Decomposition of multi-control Toffoli gates using CCNOT gates and ancillae. (a) A CCCNOT gate can be constructed using two CCNOT gates and one ancilla; (b) an $n$-CNOT gate can be constructed using $n-1$ CCNOT gates and $n-2$ ancillae. Here $c_i$ denote control qubits, $a_j$ ancillae, and $t$ the target.} \label{t1}
\end{figure}
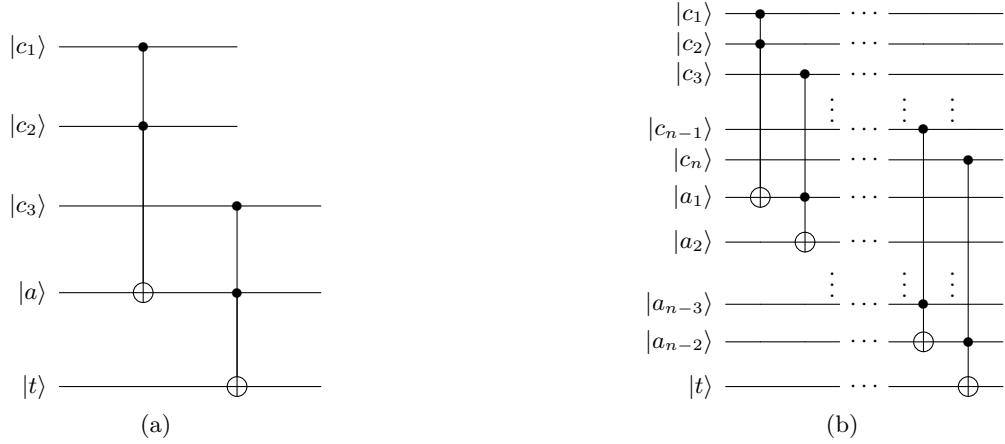

This algorithm is based on the following two points: (i) Any Boolean expression in $n$ variables with a length polynomial in $n$ can be equivalently transformed into a 3-CNF format within a number of steps polynomial in $n$~\cite{2009,Arora}, where each clause depends on at most three Boolean variables, and the number of clauses is at most polynomial in $n$; (ii) Any $n$-controlled CNOT gate ($n$CNOT) can be implemented using $n-1$ CCNOT gates and $n-2$ ancillary qubits, where an $n$CNOT gate refers to: having $n$ control bits and one target bit, such that the $X$ operation is applied to the target bit if and only if all $n$ control bits are in the state $|1\rangle$. For example, when $n=3$, two CCNOT gates and one ancillary qubit are required, as shown in Fig.~\ref{t1}(a); the general case for $n$CNOT is illustrated in Fig.~\ref{t1}(b).

Given a Boolean expression $f(x_1,x_2,\ldots,x_n)$ containing $n$ Boolean variables, according to (i) we can always transform $f$ into a 3-CNF format in time polynomial in $n$, with the number of clauses being at most polynomial in $n$, denoted as $p(n)$. We use $n$ work qubits to represent the $n$ Boolean variables, such that their $N=2^n$ possible states $|00\ldots0\rangle, |00\ldots1\rangle, \ldots, |11\ldots1\rangle$ naturally correspond to all assignments of the Boolean variables. That is, the basis vector $\ket{x}$ corresponds to the assignment $x$, where the integer $x$ is represented in binary form. Additionally, we need clause qubits $|c_1,c_2,\ldots,c_{p(n)}\rangle$ equal in number to the clauses, as well as a certain number of ancillary qubits (at most $2p(n)-2$, also polynomial in $n$), and one oracle qubit. Except for some ancillary qubits permanently set to $|1\rangle$ (which assist in implementing $X$/$Z$ and CNOT gates), all qubits start in $|0\rangle$. Following this, each work qubit undergoes a Hadamard gate $H$.

\begin{figure}[htbp]
\centering \hspace{0.0cm}
\Qcircuit @C=2em @R=2em {
    & \lstick{\ket{x_i}} & \qw & \ctrl{3} & \qw & \qw \\
    & \lstick{\ket{x_j}} & \gate{X} & \ctrl{2} & \gate{X} & \qw \\
    & \lstick{\ket{x_k}} & \gate{X} & \ctrl{1} & \gate{X} & \qw \\
    & \lstick{\ket{c_m}} & \qw & \gate{X} & \qw & \qw
}
\caption{Circuit realization of the $m$th clause $\neg x_i \vee x_j \vee x_k$ for a Boolean function $f$ expressed in 3-CNF.}
\label{fig:circuit}
\end{figure}
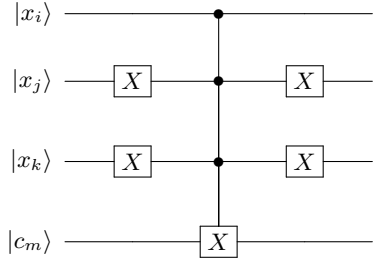

For each clause (containing at most three distinct Boolean variables), suppose the $m$th clause involves the Boolean variables $x_i, x_j, x_k$, and the clause contains only negation ($\neg$) and disjunction ($\vee$). For this clause, we can incorporate a CCCNOT gate—constructed from two CCNOT gates and one ancillary qubit—along with up to six $X$ gates (each realizable via a CCNOT gate and two auxiliary qubits permanently in the state $|1\rangle$). Specifically: if the variable $x_j$ is negated in the clause, no $X$ gate is applied to the qubit $|x_j\rangle$; if it is not negated, an $X$ gate is applied to the qubit $|x_j\rangle$. After applying all such $X$ gates, a CCCNOT gate is applied, with these three qubits as controls and the corresponding clause qubit $c_m$ as the target; subsequently, $X$ gates are applied again to $x_i, x_j, x_k$: if a qubit was acted upon by an $X$ gate before the CCCNOT, it is acted upon again by an $X$ gate at this stage; otherwise, it is not. For example, the circuit for the clause $\neg x_i \vee x_j \vee x_k$ is shown in Fig.~\ref{fig:circuit}. Processing all clauses in this manner requires at most $p(n)$ ancillary qubits.

After completing the CCCNOT gates for all clauses, we apply a $p(n)$-controlled CNOT gate with all clause qubits as controls and the oracle qubit as the target. This requires $p(n)-1$ CCNOT gates and $p(n)-2$ ancillary qubits. Clearly, after this step, for Boolean variable assignments that yield $f=1$, the corresponding oracle qubit is $|1\rangle$; for assignments yielding $f=0$, the oracle qubit is $|0\rangle$. We have thus implemented the oracle operation using only CCNOT gates, $H$ gates, and some ancillary qubits. In the following, we will represent this operation simply as a box labeled ``oracle'', and the states of the ancillary qubits and clause qubits will not affect the execution of subsequent circuits.

The real algorithm described in this section, which marks the cases $f=1$ and $f=0$ on the oracle qubit, will be employed in step (iii) of the concrete algorithm below. The entire circuit is represented simply by a box labeled ``oracle,'' as shown in Fig.~\ref{t9}(a).

\subsection{Detailed algorithm}

\begin{figure*}[th!]
\hspace*{0.5cm}
\flushleft{(a)}
	\centerline{
  \Qcircuit @C=0.35em @R=1.5em {
	 \lstick{\ket{0}} & \gate{H} & \qw& \qw  &   \multigate{7}{{\rm Oracle}}  & \qw&\gate{H}&\gate{X}&\multigate{6}{{\mathcal C}}&\qw&\multigate{6}{{\mathcal C}}&\qw&\qw&\qw&\qw&\multigate{6}{{\mathcal C}}&\qw&\qw&\qw&\qw&\qw&\qw&\qw&\qw&\qw&\qw&\qw&\qw&\qw&\qw&\qw&\qw&\qw\\
\lstick{\ket{0}}  &  \gate{H} & \qw & \qw &   \ghost{{\rm Oracle}}  & \qw &\gate{H}&\gate{X}&  \ghost{{\mathcal C}}&\qw&\ghost{{\mathcal C}}&\qw&\qw&\qw&\qw&\ghost{{\mathcal C}}&\qw&\qw&\qw&\qw&\qw&\qw&\qw&\qw&\qw&\qw&\qw\qw\qw\qw&\qw&\qw&\qw&\qw&\qw&\qw\\
 \lstick{}   &    &  &  \lstick{\vdots~~}   &    &&\rstick{\vdots~~~}& \lstick{~~~\vdots} & &&&&&&&&\lstick{\cdot\cdot~~~~~}&&&&&&&&&&\\
 %\lstick{}   &    &  &  \lstick{\vdots~~}   &    &&\rstick{\vdots~~~}& \lstick{~~~\vdots} & &&&&&&&&\lstick{\cdot\cdot~~~~~}&&&&&&&&&&\\
	 \lstick{\ket{0}}  &  \gate{H} & \qw & \qw &   \ghost{{\rm Oracle}}  & \qw &\gate{H}&\gate{X}&  \ghost{{\mathcal C}}&\qw&\ghost{{\mathcal C}}&\qw&\qw&\qw&\qw&\ghost{{\mathcal C}}&\qw&\qw&\qw&\qw&\qw&\qw&\qw&\qw&\qw&\qw&\qw\qw\qw\qw&\qw&\qw&\qw&\qw&\qw&\qw\\
\lstick{\ket{0}}  &  \qw & \qw & \qw &   \ghost{{\rm Oracle}}  & \qw &\gate{H}&\gate{X}&  \ghost{{\mathcal C}}&\qw&\ghost{{\mathcal C}}&\qw&\qw&\qw&\qw&\ghost{{\mathcal C}}&\qw&\qw&\qw&\qw&\qw&\qw&\qw&\qw&\qw&\qw&\qw\qw\qw\qw&\qw&\qw&\qw&\qw&\qw&\qw\\
\lstick{}   &    &  &  \lstick{\vdots~~}   &    &&\rstick{\vdots~~~}& \lstick{~~~\vdots} & &&&&&&&&\lstick{\cdot\cdot~~~~~}&&&&&&&&&&\\
	\lstick{\ket{0}}  &   \qw   & \qw &\qw & \ghost{{\rm Oracle}}  & \qw  &\gate{H}&\gate{X}&  \ghost{{\mathcal C}}&\qw
	 &\ghost{{\mathcal C}}&\qw  &\qw &\qw&\qw
	 &\ghost{{\mathcal C}}&\qw &\qw&\qw&\qw&\qw&\qw&\qw&\qw&\qw&\qw&\qw&\qw&\qw&\qw&\qw&\qw&\qw
	 \inputgroupv{2}{3}{4.5em}{1.1em}{{\rm work~qubits:}  ~~~~~~~~~~~~~~~~~~~~~~~~~~~~~} \inputgroupv{6}{6}{4.5em}{0.0em}{{\rm clause-ancilla:}  ~~~~~~~~~~~~~~~~~~~~~~~~~~~~~}\\
%\lstick{}   &    &  &  \lstick{\vdots~~}   &    &&\rstick{\vdots~~~}& \lstick{~~~\vdots} & &&&&&&&&\lstick{\cdot\cdot~~~~~}&&&&&&&&&&\\
	 %\lstick{\ket{0}}  &   \qw   & \qw &\qw & \ghost{{\rm Oracle}}  & \qw  &\gate{H}&\gate{X}&  \ghost{{\mathcal C}}&\qw	 &\ghost{{\mathcal C}}&\qw  &\qw &\qw&\qw	 &\ghost{{\mathcal C}}&\qw &\qw&\qw&\qw&\qw&\qw&\qw&\qw&\qw&\qw&\qw&\qw&\qw&\qw&\qw&\qw&\qw	 \inputgroupv{5}{7}{4.5em}{1.1em}{{\rm Clause~qubits and ancilla:}  ~~~~~~~~~~~~~~~~~~~~~~~~~~~~~}\\
	 \lstick{\rm oracle\ qubit : \ket{0_o}}  & \qw &\qw & \qw &   \ghost{{\rm Oracle}}  &  \qw  &\qw&\qw	 &\qw\qwx&\qw&\qw\qwx&\qw&\ustick{~\cdots}\qw&\qw&\qw&\qw\qwx&\qw&\qw&\qw&\qw&\qw&\gate{H}&\qw&\qw&\gate{X}&\qw&\ctrl{1}&\qw&\ctrl{1}&\qw&\dstick{\cdots}\qw&\ctrl{1}&\qw&\qw\\
	 \lstick{{\rm non-Hermitian\ qubit} : |0_n\rangle}  & \qw  & \qw &\qw & \qw & \qw &\qw&\qw &\gate{G}\qwx&\qw&\gate{G}\qwx &\qw
	 &\ustick{~\cdots}\qw &\qw&\qw&\gate{G}\qwx &\qw&\qw&\qw&\qw&\qw&\qw&\qw&\qw&\qw&\qw&\gate{G}&\qw&\gate{G}&\qw&\qw&\gate{G}&\qw&\qw
	 \gategroup{9}{10}{9}{15}{2.1em}{_\}}\gategroup{9}{28}{9}{31}{2.1em}{_\}}\\
	 & & &  & &  &&&&&\mbox{~~~$r$ times} &&&& &&&&&&&&&&&&&&\mbox{~~~$r'$ times}&&&\\
	 \lstick{\rm ancillary\ qubit: \ket{1}}  & \qw &\qw & \qw & \qw & \qw &\qw&\qw
	 &\qw &\qw&\qw&\qw&\qw&\qw&\qw&\qw&\qw&\qw&\qw&\qw&\qw&\qw&\qw&\qw&\ctrl{-3}&\qw&\qw&\qw&\qw&\qw&\qw&\qw&\qw \\
\lstick{\rm BHR\ qubit : \alpha\ket{0}+\beta\ket{1}}  & \qw &\qw & \qw & \qw & \qw &\qw&\qw
	 &\qw &\qw&\qw&\qw&\qw&\qw&\qw&\qw&\qw&\qw&\qw&\qw&\qw&\qw&\qw&\qw&\ctrl{-1}&\qw&\qw&\qw&\qw&\qw&\qw&\qw&\qw
	   }
  }
  \vspace{0.6cm}
\flushleft{(b)}
  \centerline{
  \Qcircuit @C=0.35em @R=1.5em {
   \lstick{} &\qw & \ctrl{4}& \qw & \qw & \qw & \qw  & \qw & \qw & \qw & & &  & \qw&\multigate{3}{{\mathcal C}}&\qw&\qw\\
  \lstick{} &\qw & \qw & \ctrl{3} & \qw &\ustick{\cdots } \qw & \qw & \qw &  \qw & \qw & & & &\qw& \ghost{{\mathcal C}}&\qw&\qw\\
  \lstick{} &    & \lstick{\vdots} & &   & & \lstick{\vdots } & \lstick{\vdots } & &  &\push{\rule{.3em}{0em}=\rule{.3em}{0em}}  &  & & & &&\\
  \lstick{} &\qw &\qw & \qw &\qw & \dstick{\cdots } \qw&\qw&  \ctrl{1}& \qw  &&  & & &\qw&\ghost{{\mathcal C}}&\qw&\qw
	 \inputgroupv{1}{4}{0.6em}{3.0em}{{\rm work\ and\ clause-ancilla\ bits:} ~~~~~~~~~~~~~~~~~~~~~~~~~~~~~~~~~~~~}\\
	 \lstick{{\rm non-Hermitian\ qubit} :}  & \qw & \gate{G} &\gate{G} & \qw  &\qw     &\qw & \gate{G}  & \qw  &  \qw &&&  &\qw&\gate{G}\qwx&\qw\qw
	 \gategroup{5}{4}{5}{7}{1.6em}{_\}}\\
	 & & &  \mbox{~~~$n$ times} &&&& &  & &&&&&&
  }
  }        \vspace{0.6cm}
\flushleft{(c)}
\centerline{
  ~~~~~\Qcircuit @C=1.2em @R=1.5em {
				\lstick{|\psi\rangle} & \qw & \gate{X} &  \ctrl{1}     &    \qw       &      \ctrl{1} &   \gate{X} &\qw    \\
				 \lstick{|\phi\rangle} & \qw & \qw &   \gate{X}&  \gate{G}   &   \gate{X}       &  \gate{G}	&\qw		
} } \hspace*{2.5cm}\vspace{20pt}
	\caption{(a) Circuit of the NQC algorithm for solving MAJSAT (a PP-complete problem). The auxiliary qubit is initialized in $\alpha|0\rangle + \beta|1\rangle$ with $\beta/\alpha = 2^i$, where $i$ is an integer ranging from $-n$ to $n$. In the figure, for simplicity, we directly depict the $X$ gates, omitting the ancillary qubits (always in $|1\rangle$) and the corresponding CCNOT gates used to realize them. (b) Detailed subcircuit $\mathcal{C}$ omitted in (a): a sequence of controlled-$G$ gates (C-$G$), each targeting the non-Hermitian qubit and controlled by a work qubit. (c) Simple implementation of C-$G$ using $X$, CNOT, and $G$, where $X$ and CNOT can be implemented by CCNOT and ancilla bits.
}
\label{t9}
\end{figure*}
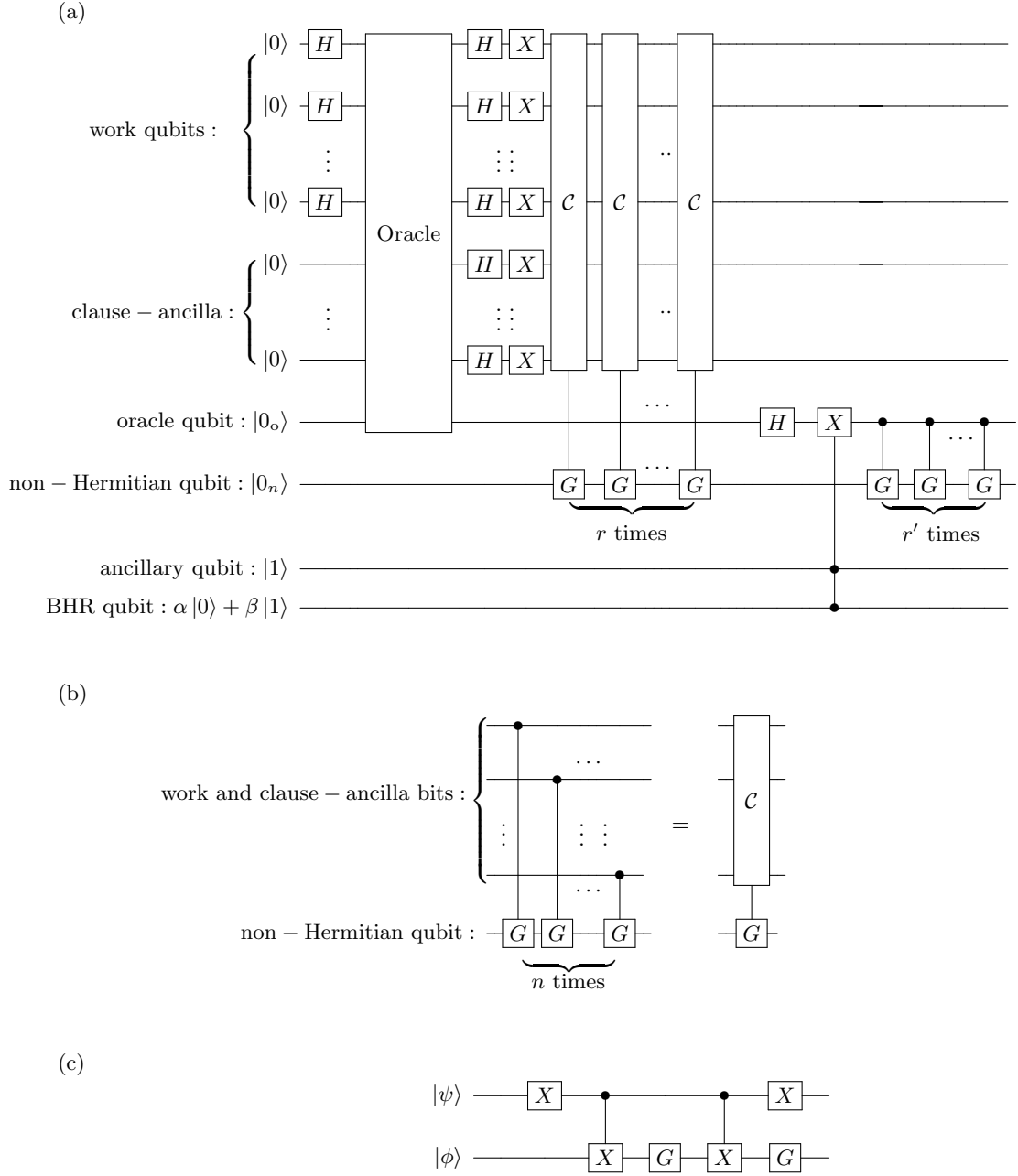

The concrete algorithmic steps as shown in Fig.~\ref{t9} are as follows (suppose $g>1$):

(i) Initialization. All qubits are initialized to $|0\rangle$ except for the Bloch-Hemi-Rotation (BHR) control qubit, which is prepared in the state
\be
\ket{\varphi_{\beta/\alpha}}=\alpha|0\rangle+\beta|1\rangle,
\ee
where $\alpha$ and $\beta$ are real and positive. The overall initial state is
\be
|\Psi_{\rm i}\rangle=|00\ldots0_w\rangle\otimes|00\ldots0_{ca}\rangle\otimes|0_o\rangle\otimes|0_n\rangle\otimes\ket{\varphi_{\beta/\alpha}},
\ee
with subscripts denoting:

$\text{w}$: work qubits,

$\text{ca}$: clause and ancillary qubits,

$\text{o}$: oracle qubit,

$\text{n}$: non-Hermitian qubit (for non-unitary transformations),

(no subscript): BHR control qubit.

Through a sequence of operations involving the BHR control qubit and the oracle qubit, the oracle qubit undergoes a rotation confined to a hemisphere of the Bloch sphere. The rotation angle is determined by the ratio $\beta/\alpha$ encoded in the initial state of the BHR control qubit. The resulting state of the oracle qubit is then recorded onto the BHR control qubit.

(ii) Apply Hadamard gates to each of the work qubits,
\be
|\Psi_{\rm ii}\rangle=\frac{1}{\sqrt{N}}\left(\sum_{j=0}^{N-1}|j_w\rangle\right)\otimes|00\ldots0_{ca}\rangle\otimes|0_o\rangle\otimes|0_n\rangle\otimes\ket{\varphi_{\beta/\alpha}}.
\ee

(iii) Apply the oracle operator $O$  to the state vector,
\begin{equation} \label{Oracle}
O=(I-P_s)\otimes I_o+P_s\otimes (|0_o\rangle\langle1_o|+|1_o\rangle\langle 0_o|)\,,
\end{equation}
where $I_o$ is the identity matrix for the oracle qubit and $P_s$ is a projection onto the sub-Hilbert space spanned by all possible solutions $|j\rangle$ of $f=1$
\be
P_s=\sum_{j\in \{f=1\}  }|j\rangle\langle j|\,.
\ee
the entire system becomes
\be
|\Psi_{\rm iii}\rangle=\frac{1}{\sqrt{N}}\left(\sum_{x=0}^{N-1} \ket{x_w}\otimes \ket{\psi(x)_{ca}}\right)\otimes
\big(a_x\ket{0_o}+b_x\ket{1_o}\big)\otimes |0_n\rangle
\otimes \ket{\varphi_{\beta/\alpha}}\, ,
\ee
where $a_x=1$ and $b_x=0$ when $f(x_1,x_2,\cdots,x_n)=0$, and $a_x=0$ and $b_x=1$ when $f(x_1,x_2,\cdots,x_n)=1$. In the combined clause and ancillary state $\ket{\psi(x)_{ca}}$, the state of each clause qubit (e.g., the $m$th) depends on the corresponding assignment of the work qubits $\ket{x_w}$. After the computation, all ancillary qubits are left in the state $\ket{1}$. This procedure has been thoroughly described in the last subsection of ``preliminaries."

(iv) Repeated application of the $\mathcal{C}$ operation. Hadamard and $X$ gates are applied to each work qubit, followed by $r$ repetitions of the $\mathcal{C}$ operation (illustrated in Fig.~\ref{t9}(b)). With $r \approx \ln N / \ln g$, this sequence of $Q$ operations effectively amplifies the $\ket{11\cdots 1}$ component among all $N = 2^n$ basis states. Neglecting terms whose coefficients are exponentially small (assuming $g > 1$), the resulting state can be approximated as
\begin{equation} \label{7}
|\Psi_{\mathrm{iv}}\rangle \approx \ket{11\cdots 1_{wca}} \otimes |\phi_o\rangle \otimes |0_n\rangle \otimes \ket{\varphi_{\beta/\alpha}},
\end{equation}
where
\begin{equation}
|\phi_o\rangle = \frac{(N-s)|0_o\rangle + s|1_o\rangle}{\sqrt{(N-s)^2 + s^2}}
\end{equation}
and
\begin{equation}
\ket{11\cdots 1_{wca}} \equiv \ket{11\cdots 1_w} \otimes \ket{11\cdots 1_{ca}}.
\end{equation}

At this stage, the number of satisfying assignments $s$ is encoded in the coefficients of the oracle qubit state $|\phi_o\rangle$. Since the difference between these two coefficients can be exponentially small, distinguishing them generally requires an exponential number of measurements.

(v) Apply a Hadamard gate to the oracle qubit, followed by a CNOT gate controlled by the BHR qubit with the oracle qubit as the target. Here we make use of a CNOT gate, which can be implemented using a CCNOT gate and an ancillary qubit. In this construction, the ancillary qubit serves as one of the control bits of the CCNOT gate, and its state remains $|1\rangle$ from beginning to end. %For simplicity, this ancillary qubit is omitted in Fig.~\ref{t9}.
The resulting state is approximately
\be
|\Psi_{\rm v}\rangle\approx\frac{|11\ldots1_{wca}\rangle\otimes\left\{\alpha \left[N|0_o\rangle+(N-2s)|1_o\rangle\right]\otimes|0_n\rangle\otimes|0\rangle+\beta\left[(N-2s)|0_o\rangle+N|1_o\rangle\right]\otimes|0_n\rangle\otimes|1\rangle\right\}}{\sqrt{2(N-2s)^2+2N^2}}.
\ee

(vi) With the oracle qubit as the control and the non‑Hermitian qubit as the target, apply the C‑$G$ gates $r'$ times. By choosing $r'\approx \ln N/\ln g$, a measurement of the oracle qubit will almost certainly yield $\ket{1_o}$, and the BHR qubit ends up in the state
\begin{equation} \label{PPS}
|\varphi_{\beta/\alpha}'\rangle\approx\frac{\alpha(N-2s)|0\rangle+\beta N|1\rangle}{\sqrt{\alpha^2(N-2s)^2+\beta^2 N^2}},.
\end{equation}
The reason why Eqs.~(\ref{7}) and (\ref{PPS}) are valid approximations when $r\approx\ln N/\ln g$ and $r'\approx\ln N/\ln g$—i.e., the total amplitude of all discarded terms is exponentially small in $n$—can be found in the Lorentz quantum computer algorithm for MAJSAT presented in \cite{ZhangWu}. Although the explicit expressions differ, the order of magnitude analysis is the same.

(vii) Measure the BHR qubit in the $x$ direction a large number of times and count the occurrences of the two outcomes $+1$ and $-1$. Concretely, we perform $n$ sets of identical circuits as shown in Fig.~\ref{t9}, each set consisting of a number of runs proportional to $n$. In every run the BHR qubit is measured in the $x$ basis. For a given set, if the number of $-1$ outcomes exceeds that of $+1$ outcomes, the set is labeled “success”; otherwise it is labeled “failure”. We need to determine whether all $n$ sets are successful. This requires repeating steps (i)–(vi) a total of $O(n^2)$ times. The basic idea is the same as in \cite{ZhangWu}.

(viii) Repeat the whole procedure $2n+1$ times, each time with a distinct value of $\beta/\alpha = 2^i$ for the BHR qubit, where $i$ is an integer ranging from $-n$ to $n$ inclusive. If $s > 2^{N-1}$—i.e., the input $\omega$ (the Boolean expression $f$ in $n$ variables) encoded in the oracle belongs to PP—then there exists at least one choice of $\beta/\alpha = 2^i$ (with $i$ in that range) for which all $n$ sets of measurements are successful. Conversely, if $s \le 2^{N-1}$, then for every such $\beta/\alpha$, all $n$ sets result in failure. In this way we solve MAJSAT, a PP-complete problem. A detailed justification of this criterion can be found in Ref.~\cite{ZhangWu}.

\subsection{${\text P}^{\sharp \text{P}}\subseteq$BRNQP}

In computational complexity theory, it is known that $\text{P}^{\text{PP}} = \text{P}^{\sharp\text{P}}$. As a consequence, any $\mathbb{R}$ NQC algorithm capable of solving PP problems efficiently can be adapted to solve problems in $\sharp\text{P}$ and $\text{P}^{\sharp\text{P}}$. A concrete demonstration of this principle using a Lorentzian-type quantum computer was provided in Ref.~\cite{ZhangWu}, and the same reasoning applies in the present context.

Bounded-error complexity classes—such as $\text{BQP}$, $\text{BNQP}$, and those involving $\text{PostBQP}$ oracles~\cite{aaronson}—are inherently probabilistic. In each such computation, the probability of error is non-negligible, but bounded below $1/2$ by at least an inverse polynomial. Through polynomial-time repetition, this error can be reduced to exponentially small, which underlies the robustness of bounded-error computation. Moreover, when all operations are unitary and the circuit size is polynomially bounded, an exponentially small error remains exponentially small throughout the computation. This explains why $\text{BQP}^{\text{BQP}} = \text{BQP}$ holds~\cite{Bernstein}.

The situation changes fundamentally when non-unitary operations or postselection are introduced. In these settings, an exponentially small error—though harmless in the purely unitary regime—can be catastrophically amplified by subsequent postselection or non-unitary gates. Consequently, adaptive oracle calls become problematic when the oracle itself is equipped with such enhanced operations. As a result, we do not obtain identities such as $\text{BNQP} = \text{BNQP}^{\text{BNQP}}$ or $\text{BRNQP} = \text{BRNQP}^{\text{BRNQP}}$.

To maintain control over error propagation, any oracle call involving postselection or non-unitarity must be made \emph{non-adaptively}. That is, all queries must be submitted in parallel, and later queries may not depend on the outcomes of earlier ones. Additionally, the main circuit itself must contain no non-unitary gates or postselection. Under these constraints, we obtain the relation $\text{BNQP} = \text{BQP}^{\text{BNQP}}_{\parallel}$.

In contrast, if the main circuit is classical, it cannot process quantum superpositions; it can only handle definite computational basis states. Each oracle call then effectively performs a measurement of the quantum state, during which any exponentially small errors are eliminated. This allows oracle calls to be made adaptively. Moreover, since the main circuit runs in polynomial time, the number of oracle calls is polynomially bounded, and each contributes only an exponentially small error. Hence, the total error can be rendered negligible. This yields $\text{P}^{\text{BNQP}} = \text{BNQP}$ and $\text{P}^{\text{BRNQP}} = \text{BRNQP}$. In Ref.~\cite{ZhangWu}, we introduced the Lorentz quantum computer and defined the associated complexity class $\text{BLQP}$, establishing that $\text{BLQP} = \text{P}^{\text{BLQP}}$. The results for $\text{BNQP}$ and $\text{BRNQP}$ are analogous~\cite{ZhangWu}.

Since we have solved a $\text{PP}$-complete problem in polynomial time using $\mathbb{R}$ NQC—implying $\text{PP} \subseteq \text{BRNQP}$—it follows that
\be
\text{P}^{\sharp\text{P}} = \text{P}^{\text{PP}} \subseteq \text{P}^{\text{BRNQP}} = \text{BRNQP}.
\ee

\section{BNQP $\subseteq$ \pnp}

Following the proof of BQP $\subseteq$ \pnp~\cite{Bernstein} with appropriate modifications, we now establish that BNQP $\subseteq$ \pnp. Consider a polynomial-size non-Hermitian quantum circuit consisting of a sequence of logical gates $N_1, N_2, \ldots, N_t$, where $t = p(n)$ for some polynomial function $p$ of the input length $n$. Since the algorithm runs in polynomial time, the number of gates is at most polynomial. Such a non-Hermitian circuit can be represented as
\be
|f\rangle=|N_tN_{t-1}\ldots N_1|i\rangle,
\ee
where $|i\rangle$ denotes the input state and $|f\rangle$ the output state.

According to the definition of BNQP~\cite{zhangwu}, a language L belongs to BNQP if and only if there exists a uniform family of non-Hermitian quantum circuits $\{\mathbb{C}_n\}_{n\geq1}$ such that for any instance $\omega$ of length $q(n)$ (where $q(n)$ is a monotonically increasing polynomial function), the output state contains a designated ``yes'' qubit $|\phi_Y\rangle$ whose state indicates the outcome. Specifically, if $\omega \in $ L, the ``yes'' qubit is in state $|1_Y\rangle$ with probability at least $2/3$; otherwise, it is in $|1_Y\rangle$ with probability at most $1/3$ (equivalently, in $|0_Y\rangle$ with probability at least $2/3$). In this bounded-error setting, the error can be reduced exponentially by repeating the algorithm a polynomial number of times. To compute the success probability, we introduce the projector (for circuit $\mathbb{C}_n$, i.e., for instances of length $q(n)$)
\be \label{YES}
P_{\text{yes}}=\left[\sum_{j}|j_1\rangle\langle j_1|\otimes\ldots\otimes\sum_{j}|j_s\rangle\langle j_s|\right]\otimes |1_Y\rangle\langle 1_Y| ,
\ee
where $|j_k\rangle$ denotes the state of the $k$th qubit in the computational basis ($j = 0$ or $1$), and the total number of qubits $s$ is at most polynomial in $n$. With this operator, the probability that the ``yes'' qubit is measured in $|1_Y\rangle$ is given by
\be \label{01}
|c_{\text{yes}}|^2=\langle i| N_1^\dag N_2^\dag \ldots N_t^\dag|P_{\text{yes}}|N_tN_{t-1}\ldots N_1|i\rangle.
\ee

To relate this to \pnp, we expand the expression as
\ba \label{1} \nonumber
|c_{\text{yes}}|^2&=&\sum_{z_1,z_2,\ldots,z_{2t}} \langle i|N_1^\dag|z_{2t}\rangle\langle z_{2t}|N_2^\dag|z_{2t-1}\rangle\ldots \langle z_{t+2}|N_t^\dag|z_{t+1}\rangle\langle z_{t+1}| P_{\text{yes}}|z_{t}\rangle \\ && \langle z_{t}|N_t|z_{t-1}\rangle   \ldots \langle z_{2}|N_2|z_{1}\rangle\langle z_{1}|N_1|i\rangle,
\ea
where each $|z_i\rangle$ represents a complete orthonormal basis vector in the computational basis. Equation~\eqref{1} sums over $2^{2p(n)}$ terms, which is exponential in $n$. From this, we define a language in P as follows. Consider a deterministic Turing machine, called the $+$DTM, whose inputs are tuples of the form $(|z_1\rangle, \ldots, |z_{2t}\rangle, |i\rangle, N_1, \ldots, N_t, k)$, where $0 \le k \le 2^{nc}$ is a real number with minimum spacing $2^{-nc}$. The total input length is linear in $p(n)$ and thus polynomial in $n$, as $p(n)$ itself is polynomial. For such an input, the $+$DTM computes
\be
a=\langle i|N_1^\dag|z_{2t}\rangle\langle z_{2t}|N_2^\dag|z_{2t-1}\rangle\ldots \langle z_{t+2}|N_t^\dag|z_{t+1}\rangle\langle z_{t+1}| P_{\text{yes}}|z_{t}\rangle \langle z_{t}|N_t|z_{t-1}\rangle   \ldots \langle z_{2}|N_2|z_{1}\rangle\langle z_{1}|N_1|i\rangle.
\ee
If $\operatorname{Re}(a) > 0$ and $k < \operatorname{Re}(a)$, the $+$DTM outputs $M = 1$ (accept); otherwise, it outputs $M = 0$ (reject). Each gate $N_i$ acts on only a few qubits. Although $P_{\text{yes}}$ in Eq.~\eqref{YES} contains exponentially many elements in the computational basis, computing $\langle z_{t+1}| P_{\text{yes}} |z_t\rangle$ is straightforward and takes constant time. Hence, deciding the output of the $+$DTM for a given input is a problem in P relative to the input length $n$ of the original NQC. However, the number of possible instances for the $+$DTM is exponential.

Next, we introduce a $\sharp$P oracle. This oracle, given a P-language and an exponential list of inputs, returns the number of inputs for which the output is $M = 1$. Feeding all possible tuples $(|z_1\rangle, \ldots, |z_{2t}\rangle, |i\rangle, N_1, \ldots, N_t, k)$ to the $\sharp$P oracle yields exactly the number of terms contributing positively to Eq.~\eqref{1}, multiplied by $2^{nc}$. Thus, with the help of the $\sharp$P oracle, we directly obtain the positive part of Eq.~\eqref{1}. Similarly, we define a $-$DTM with inputs of the same form but with $-2^{nc} \le k \le 0$, which accepts if $\operatorname{Re}(a) < 0$ and $k > \operatorname{Re}(a)$. Using another $\sharp$P oracle, we obtain the negative part of Eq.~\eqref{1}. In this way, with two calls to a $\sharp$P oracle, we can compute $|c_{\text{yes}}|^2$.

A remark is in order. In defining the $+$DTM and $-$DTM, we introduced a constant $c$ that controls the precision: the larger the $c$, the finer the spacing of $k$. As $n$ increases, the spacing between consecutive $k$ values decreases exponentially, which is exactly what we need.

Since the gates are not required to be unitary, the final state may not be normalized. To obtain the success probability, we also need the norm of the final state. This can be done analogously by replacing $P_{\text{yes}}$ with
\be \label{YN}
P_{\text{yn}}=\left[\sum_{j}|j_1\rangle\langle j_1|\otimes\ldots\otimes\sum_{j}|j_s\rangle\langle j_s|\right]\otimes \left[|0_Y\rangle\langle 0_Y|+|1_Y\rangle\langle 1_Y|\right],
\ee
and following the same procedure to compute $|c_{\text{yes}}|^2 + |c_{\text{no}}|^2$ using two further calls to the $\sharp$P oracle. The acceptance probability of the NQC circuit is then $\frac{|c_{\text{yes}}|^2}{|c_{\text{yes}}|^2 + |c_{\text{no}}|^2}$. Consequently, any language L decidable by a polynomial-time NQC circuit—i.e., L $\in \text{BNQP}$—can be decided with just four calls to a $\sharp$P oracle, implying $\text{BNQP} \subseteq $ \pnp.

Since we have in Sec.~II shown that $\text{P}^{\sharp\text{P}} \subseteq \text{BRNQP}$ and it is clear that $\text{BRNQP} \subseteq \text{BNQP}$ (as BRNQP corresponds to a restricted version of BNQP), we obtain the inclusion chain
\be
\text{BNQP} \subseteq {\text P}^{\sharp \text{P}} \subseteq \text{BRNQP}\subseteq \text{BNQP}
\ee
which implies $\text{BRNQP} = \text{BNQP} = \text{P}^{\sharp\text{P}}$. It is important to note that while we have shown that any single NQC circuit deciding a language can be simulated with only four $\sharp \text{P}$ oracle queries, solving a \pnp-complete problem generally requires a polynomial number of such circuits and measurements. Hence, the equality $\text{BRNQP} = \text{BNQP} = \text{P}^{\sharp\text{P}}$ is consistent.

\section{Summary}

In this work, we prove that $\text{BRNQP} = \text{BNQP} = \text{P}^{\sharp\text{P}}$, where $\text{BRNQP}$ is defined by polynomial-time bounded-error circuits over the real non-unitary gate set ${H, \text{CCNOT}, G=\operatorname{diag}(g^{-1},g)}$ (with $g>0$, $g\neq1$) and $\text{BNQP}$ is its universal counterpart. The proof proceeds in two directions: $\text{P}^{\sharp\text{P}} \subseteq \text{BRNQP}$ via an explicit construction, and $\text{BNQP} \subseteq \text{P}^{\sharp\text{P}}$ by adapting the classical proof of $\text{BQP} \subseteq \text{P}^{\sharp\text{P}}$ to the non-Hermitian setting.

A key consequence is that this real non-unitary model, despite lacking gate universality, achieves the full computational power of its universal non-Hermitian counterpart. Our result reveals that even a minimal non-unitary operation elevates quantum computing to the level of $\text{P}^{\sharp\text{P}}$—far beyond standard quantum computing, which is believed to be contained in $\text{BQP}$—and demonstrates that complex numbers are unnecessary for this power, as a purely real gate set suffices. Overall, our work bridges non-unitary quantum computing with classical complexity theory and provides new insights into the roles of non-unitarity and reality as computational resources.

\section*{ACKNOWLEDGMENTS}
I would like to thank Biao Wu from Peking University for insightful discussions. This work was supported by the National Natural Science Foundation of China (Grant No. 12475018) and the Shanghai Municipal Science and Technology Project (Grant No. 25LZ2601100).

%\appendix

\end{CJK*}
\end{document}